\RequirePackage[2020-02-02]{latexrelease}
\documentclass[onecolumn,showpacs,preprintnumbers,amsmath,amssymb,superscriptaddress]{revtex4}
\usepackage{graphicx}
\usepackage{dcolumn}
\usepackage{bm}
\usepackage{tabularx}
\usepackage{ulem}
\usepackage{amsmath,amssymb} 
\newcolumntype{M}{>{\centering\arraybackslash}m{1.85cm}}
\usepackage[export]{adjustbox}
\usepackage{float}
\usepackage{braket}
\usepackage{makecell}
\usepackage{lipsum}
\usepackage{longtable}
\graphicspath{{figure/}}
\newcommand\T{\rule{0pt}{3ex}}       
\newcommand\B{\rule[-1.5ex]{0pt}{0pt}} 
\usepackage{xcolor}   

\makeatletter
\newcommand{\colorcaption}[2][]{%
	\begingroup%
	\renewcommand{\@caption@fignum@sep}{ (Color online). }%
	\caption[#1]{#2}%
	\endgroup%
}
\makeatother
\begin{document}

	\title{Systematic shell-model study of $^{98-130}$Cd isotopes and $8^+$ isomeric states}
	
	\author{ Deepak Patel$^{1}$, Praveen C. Srivastava\footnote{Corresponding author: praveen.srivastava@ph.iitr.ac.in}}
	\address{Department of Physics, Indian Institute of Technology Roorkee, Roorkee 247667, India}
	\author{Noritaka Shimizu}
	\address{Center for Computational Sciences, University of Tsukuba, 1-1-1, Tennodai Tsukuba, Ibaraki 305-8577, Japan}
	
	\date{\hfill \today}
	
	
	\begin{abstract}
 We present systematic shell-model studies of even-even $^{98-130}$Cd isotopes using a realistic effective shell-model interaction derived from the G-matrix approach with an inert core $^{88}$Sr. Our calculated low-lying excited energy spectra and electromagnetic properties are compared with the experimental data. On the basis of recently available experimental data, we predict spins and parities corresponding to unconfirmed states. We also discuss the properties of $8^+$ isomeric states in $^{98-104,130}$Cd isotopes. 	
	\end{abstract}
	
	
	\maketitle
	
	
	\section{Introduction}
	
In recent years, the study of nuclear structure properties of Cd and Te isotopes has increased both from the experimental and theoretical sides because they are in the vicinity of the  $^{100}$Sn, which is the heaviest $N=Z$ doubly magic nucleus \cite{Siciliano, Vaman, Togashi, Nara Singh, Faestermann}. 
 The study of nuclear properties in Cd isotopes is also essential for the understanding of stellar nucleosynthesis. Several similarities are pointed out between Sn and Cd isotopic chains, like excitation energy spectra and electric quadrupole transition probability \cite{Togashi, Zuker, Banu, Doornenbal, Isacker, Jain, Rakesh, Srivastava}. The shell model provides a better platform to verify these properties.

	In the past, the structures of low-lying states of Cd-isotopes have been investigated using different advanced experimental techniques. For the first time, Alber $et$ $al.$ \cite{Alber} experimentally observed the low-lying spectra of neutron-deficient $^{100-102}$Cd. Measurements of lifetimes of excited nuclear states are important for computing $B(E2)$ transition rates. Lieb $et$ $al.$ \cite{Lieb} determined picosecond lifetimes of high-spin states in $^{102}$Cd isotope using the reaction $^{58}$Ni($^{50}$Cr,$\alpha$ $2p$) and coincidence recoil distance Doppler-shift technique (RDDS). Boelaert $et$ $al.$ \cite{Boelaert1} determined the lifetimes of low-lying $2^{+}_{1}$ and $4^{+}_{1}$ states in $^{104}$Cd and $2^{+}_{1}$ state in $^{102}$Cd. M\"uller $et$ $al.$ \cite{Muller} determined lifetimes of high-spin states in $^{104}$Cd in the time range 0.3 $ps$$\le\tau\le$1.2 $ns$ for the first time using RDDS and Doppler-shift attenuation methods (DSAM). Simons $et$ $al.$ \cite{Simons} measured the lifetimes of excited states in the lowest-lying positive parity bands in $^{106,108}$Cd isotopes using the RDDS and also deduced the $B(E2)$ transition rates. Benczer-Koller $et$ $al.$ \cite{Koller} remeasured and supplemented existing lifetimes and magnetic moments of low-lying states in $^{106}$Cd. Fan $et$ $al.$ \cite{Fan} measured g-factors of rotational states of the yrast band from spin $10^{+}$ to spin $16^{+}$ in $^{108}$Cd using transient-magnetic-field ion implantation perturbed angular distribution method. Kumpulainen $et$ $al.$ \cite{Kumpulainen} investigated the low-lying low-spin collective states in even $^{106-112,116}$Cd isotopes using in-beam and off-beam $\gamma$-ray and conversion-electron spectroscopy. Gade $et$ $al.$ \cite{Gade1} reported the first evidence for the complete quadrupole-octupole coupled quintuplet of negative parity $(2^{-}_1\otimes3^{-}_1)^{(J^-)}$ with $J=1,2,3,4,5$ for $^{108}$Cd. 
    Corminboeuf $et$ $al.$ \cite{Corminboeuf} performed measurements consisting of $\gamma$-ray excitation functions and angular distributions on $^{110}$Cd. They extracted the lifetimes for 16 excited states using the DSAM. Piiparinen $et$ $al.$ \cite{Piiparinen} measured lifetimes of 20 yrast levels in $^{110}$Cd using the recoil-distance method and the NORDBALL array of Compton-suppressed Ge-detectors. They also deduced the reduced-transition probabilities and revealed the rotational and vibrational behavior of $^{110}$Cd. Garrett $et$ $al.$ \cite{Garrett} measured lifetimes of those levels which are below 4 MeV in $^{112}$Cd isotope using DSAM following the inelastic scattering of monoenergetic neutrons. Jamieson $et$ $al.$ \cite{Jamieson} studied the quadrupole-octupole coupled states in $^{112}$Cd with $^{111}$Cd$(d,p)^{112}$Cd reaction using 22 MeV polarized deuterons. Zamfir $et$ $al.$ \cite{Zamfir} studied the structure of $^{122}$Cd through the decay of $^{122}$Ag by $\gamma-\gamma$ coincidences, $\gamma$-ray angular correlations, and $\beta-\gamma-\gamma$ fast electronic scintillation timing (FEST) measurements. Scherillo $et$ $al.$ \cite{Scherillo} investigated microsecond isomers in Cd isotopes in the mass range A=123 to 130 using the LOHENGRIN mass spectrometer. Hoteling $et$ $al.$ \cite{Hoteling} identified the presence of isomeric levels with half-lives in the microsecond range in $^{125-128}$Cd. In a recent TITAN experiment at TRIUMF, the cadmium masses were measured for $^{125/m}$Cd and $^{126}$Cd. Lascar $et$ $al.$ \cite{Lascar} presented the results of precision mass measurements of $^{125-127}$Cd isotopes. They have also compared their measurements with the $ab$ $initio$ shell-model calculations. Dunlop $et$ $al.$ \cite{Dunlop} measured the $\beta$-decay half-lives of $^{128-130}$Cd isotopes using GRIFFIN $\gamma$-ray spectrometer at the TRIUMF-ISAC facility. Dillmann $et$ $al.$ \cite{Dillmann} performed the first $\beta$- and $\gamma$-spectroscopic decay of the $N=82$ r-process ``waiting-point" nuclide $^{130}$Cd at CERN-ISOLDE.
	
	In the previous theoretical works, Schmidt $et$ $al.$ \cite{Schmidt} performed shell-model calculations for even-even $^{98-108}$Cd isotopes.
 Nomura $et$ $al.$ \cite{Nomura} investigated the structure of even-even $^{108-116}$Cd isotopes by combining the self-consistent mean-field approach and the interacting boson model.
    Blazhev $et$ $al.$ \cite{Blazhev} identified a core excited $I^{\pi}$=(12$^{+}$) spin-gap isomer in $^{98}$Cd. They have also discussed the measured $E4$ and $E2$ strengths, $^{100}$Sn core excitations, and the origin of the empirical polarization charges in large-scale shell-model calculations.
	Boelaert $et$ $al.$ \cite{Boelaert} performed shell-model calculations for even-even $^{98-106}$Cd isotopes and calculated energy spectra and transition probabilities. Clark $et$ $al.$ \cite{Clark} studied the gamma decay of excited states in $^{100}$Cd with Gammasphere array following the $^{46}$Ti($^{58}$Ni,2$p$,2$n$) reaction at 215 MeV, which were compared with shell-model results. Maheshwari $et$ $al.$ \cite{Maheshwari} studied the energy variation of $2^{+}_{1}$ states and $B(E2;0^{+}_{1}\rightarrow2^{+}_{1})$ trends in the Cd, Sn, and Te isotopic chains using generalized seniority approach.

To study the structural evolution of the even-even Cd isotopes, we performed systematic shell-model calculations for even $^{98-130}$Cd isotopes.
 We mainly discuss the behavior of excitation energy spectra, the inverted parabolic shape of $B(E2; 0^+_1\rightarrow 2^+_1)$ transition in the Cd chain, and the isomeric $8^+$ states.  Since several new experimental data have been available in cadmium isotopes for low-lying spin states recently, we here focus on the low-lying spin states of these isotopes.
	
	This paper is divided into the following sections. In Section \ref{section2}, we have discussed briefly the interaction used in our calculations. In Section \ref{section3}, we introduce the monopole shifts in the single-particle energies of valence proton and neutron orbitals for Cd-isotopes. In Section \ref{section4}, we present our calculated results of low-lying energy spectra, electromagnetic properties, octupole collective $3^{-}$ states, and $8^{+}$ isomeric states in the Cd chain, we have also presented the calculated result for energy surface using the Hartree-Fock-Bogoliubov (HFB) method.  Finally, we summarize our results and conclude the paper in section \ref{section5}.

\section{Shell-model framework} \label{section2}
 
	The nuclear shell-model Hamiltonian can be expressed in terms of single-particle energies and two nucleon interactions as	
	\begin{equation}
	\hat{H} = \sum_{\alpha}\epsilon_{\alpha}\hat{n}_\alpha +\sum_{\alpha\leq \beta, \gamma \leq \delta, J, M}\langle j_\alpha j_\beta |V|j_\gamma j_{\delta} \rangle_{J}
  A^\dagger_{\alpha,\beta}(JM) A_{\gamma, \delta}(JM) , 
	\end{equation}
	\begin{equation}
\hat{n}_\alpha = \sum_{m_\alpha}
c^\dagger_{\alpha, m_\alpha}c_{\alpha, m_\alpha}, \ \ \ A^\dagger_{\alpha,\beta}(J,M) = 
\frac{1}{\sqrt{1+\delta_{\alpha\beta}}}
\sum_{m_\alpha,m_\beta}
\langle j_\alpha m_\alpha j_\beta m_\beta| J M \rangle c^\dagger_{\alpha, m_\alpha}c^\dagger_{\beta, m_\beta}, 
	\end{equation}
	where $\alpha=\{n,l,j,t_z\}$ stand for the single-particle orbitals and $\epsilon_{\alpha}$ denots the single-particle energy of the orbit $\alpha$. $c_{\alpha,m_\alpha}$ and $c_{\alpha,m_\alpha}^{\dagger}$ are the fermion annihilation and creation operators of the single particle state $(\alpha,m_\alpha)$. $\langle j_\alpha j_\beta |V|j_\gamma j_\delta \rangle_{J}$ 
    is  called the two-body matrix element (TBME).
    
	We study the structure of even-even Cd isotopes from the neutron numbers 50 to 82 by performing the shell-model calculations using the G-matrix effective interaction, which was constructed by the effective microscopic interaction derived from a charge-symmetry breaking nucleon-nucleon potential \cite{Machleidt}
 with further modifications in the monopole part and was used in Ref.~\cite{Boelaert}. The model space consists of the $1p_{1/2}$, $0g_{9/2}$ proton orbitals and the $1d_{5/2}$, $2s_{1/2}$, $1d_{3/2}$, $0g_{7/2}$, $0h_{11/2}$ neutron orbitals with the inert core $^{88}$Sr. The shell model code KSHELL \cite{KShell} was used for our calculations to diagonalize the Hamiltonian matrices. The largest $M$-scheme dimension in the present study is 0.96 $\times$ 10$^9$ for $^{114}$Cd.

	\section{Monopole interaction and effective single-particle energies} \label{section3}
	
	In the case of $^{98-130}$Cd, there are two valence proton holes and several valence neutrons outside the doubly magic nucleus $^{100}$Sn. It can be expected that the nucleon-nucleon interaction will influence the effective single-particle energies of the proton and neutron orbitals. The variation in the single-particle energy of valence proton and neutron orbitals as a function of neutron number is known as the monopole shift \cite{Srivastava2}. 

    The shell-model two-body interaction can be divided into the monopole part and the remaining multipole part as
    \begin{equation}		\hat{V}=\hat{v}_\textrm{mono}+\hat{v}_\textrm{multi}.
    \end{equation}
    The monopole interaction is defined as \cite{Otsuka}
\begin{equation}
\hat{v}_\textrm{mono}= \frac12 \sum_{\alpha,\beta} 
\bar{E}_{\alpha,\beta} : \hat{n}_\alpha \hat{n}_\beta:
\end{equation}
where $::$ denotes the normal ordering and the monopole matrix element is defined as
\begin{equation}
	\bar{E}(\alpha,\beta)=\frac{\sum_{J}(2J+1)
 \langle \alpha\beta|V|\alpha\beta\rangle_J}{\sum_{J}(2J+1)}.
\end{equation}

The effective single-particle energy (ESPE) of the orbit $\alpha$ is the sum of the single-particle energy with respect to the inert core and the contribution from the valence particles as
\begin{equation}
\tilde{\epsilon}_\alpha=\epsilon_\alpha+\sum_{\beta}\bar{E}(\alpha,\beta)\tilde{n}_{\beta}, 
\end{equation}
where $\tilde{n}_{\beta}$ is the occupation number of the orbit $\beta$ of the filling configuration \cite{Otsuka}.
In the case of $\alpha=\beta$ and $\alpha$ is the occupied orbit in the filling configuration, 
 $\tilde{n}_\beta$ is the occupation number minus one.

\begin{figure}
	\vspace{-0.8cm}
	\includegraphics[width=78mm]{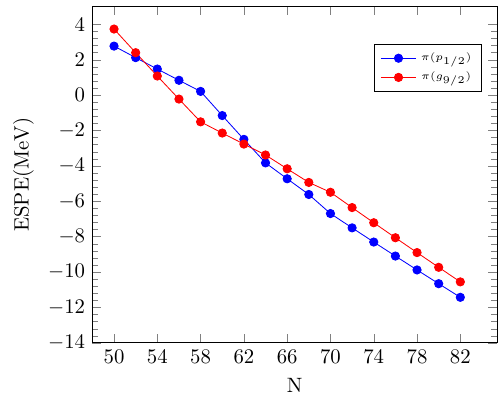}
	\includegraphics[width=79mm]{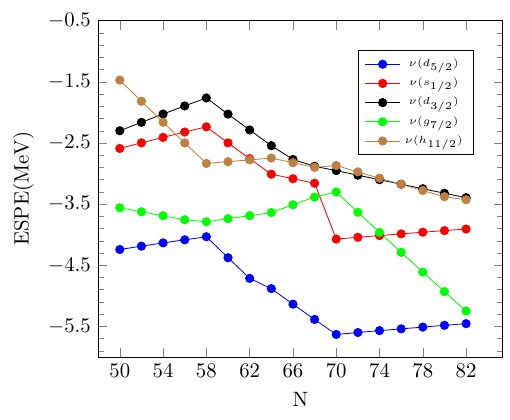}
	\caption{\label{fig1} Variation in the effective single-particle energies of proton and neutron orbitals in Cd isotopes as a function of neutron number.}
\end{figure}

Figure \ref{fig1} shows the change in effective single-particle energy (ESPE) of proton and neutron orbitals with respect to neutron number in the Cd isotopes for the G-matrix interaction. The variation of the effective single-particle energies of proton orbitals depends on the proton-neutron effective interactions \cite{Smirnova}. 
We can see that the decrement rate in the ESPE of the proton $g_{9/2}$ orbital is larger for neutron-deficient Cd isotopes than the $p_{1/2}$ orbital. For the mid-shell isotopes, the rate of variation in ESPEs is changed for both orbitals, and the ESPEs of $\pi(p_{1/2})$ decrease faster than the $\pi(g_{9/2})$. In the case of neutron-rich isotopes, the variation is almost constant.

The ESPEs of neutron orbitals vary in a different way than the ESPEs of proton orbitals. The ESPEs of neutron orbitals are not decreasing almost monotonically like proton orbitals with the increase of neutron number. It is notable that the ESPE of $\nu(d_{5/2})$ orbital lies lower than the other four neutron orbitals for the whole chain. It implies that the neutron particles will be occupied first in this orbital with more dominance.

	\section{Results} \label{section4}
	
	\begin{figure*}
		\includegraphics[width=182mm, height = 90mm]{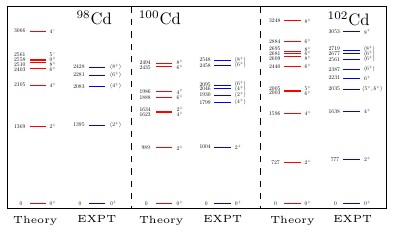}
		\includegraphics[width=182mm, height = 90mm]{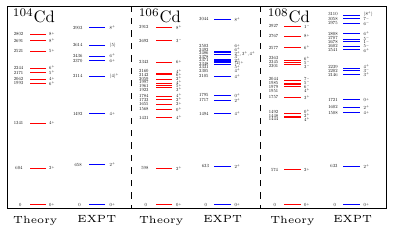}
		\caption{\label{fig2} Comparison between calculated and experimental energy \cite{NNDC} levels for even-even  $^{98-108}$Cd isotopes.}
	\end{figure*}

	\subsection{Low-lying energy spectra of $^{98-130}$Cd even isotopes }
	
\begin{figure*}
		\includegraphics[width=182mm, height = 90mm]{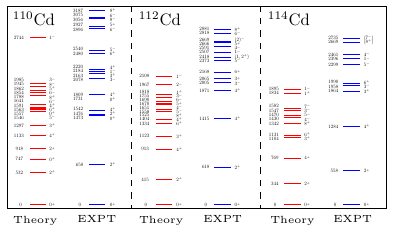}
		\includegraphics[width=182mm, height = 90mm]{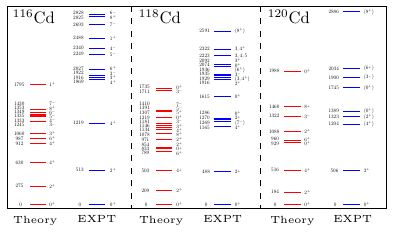}
		\caption{\label{fig3} Comparison between calculated and experimental energy \cite{NNDC} levels for even-even  $^{110-120}$Cd isotopes.}
	\end{figure*}

\begin{figure*}
		\includegraphics[width=182mm, height = 90mm]{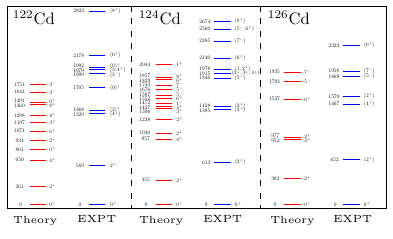}
		\includegraphics[width=182mm, height = 90mm]{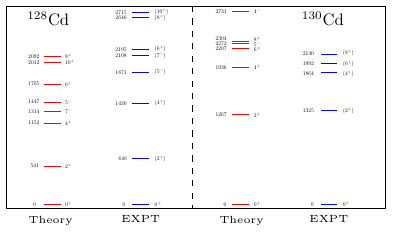}
		\caption{\label{fig4} Comparison between calculated and experimental energy \cite{NNDC} levels for even-even  $^{122-130}$Cd isotopes.}
	\end{figure*}

	We discuss the shell-model energy spectra of even Cd isotopes using the G-matrix interaction and compared them with the experimental data. The energy levels are shown in Figs. \ref{fig2}-\ref{fig4}. We have shown five states maximally for a given spin-parity $J^{\pi}$.
	
	{\bf $^{98}$Cd}: G\'orska $et$ $al.$ \cite{Gorska} identified and studied excited states in $^{98}$Cd (two proton holes from $^{100}$Sn) using in-beam spectroscopy for the first time. Our calculation shows good agreement with the experimental data for the excited $2_1^{+}$, $4_1^{+}$, $6_1^{+}$, and $8_1^{+}$ states. The ground state $0^{+}$ and other excited states up to $8^{+}$ show $\pi(g_{9/2}^{-2})$ character.
  We have also shown three other states $0^{+}_2$, $5^{-}_1$ and $4^{-}_1$ to compare the results obtained in Ref. \cite{Gorska}. 
  
	{\bf $^{100}_{48}$Cd$_{52}$}: The  $^{100}_{48}$Cd$_{52}$ isotope is a good candidate for studying two valence neutron-particles and  proton-holes configurations. The excited energies of $2^{+}_{1}$, $4^{+}_2$, $6^{+}_{2}$, and $8^{+}_{1}$ states by the shell-model calculations are showing the reasonable agreement with the experimental data \cite{NNDC}. The $4^{+}_1$ and $6^{+}_1$ states lie slightly lower than the tentative experimental data with our calculation. The shell-model result predicts the configurations of $0^{+}_1$, $2^{+}_1$ and $4^{+}_1$ state as $\pi(p_{1/2}^2g_{9/2}^8)\otimes\nu(d_{5/2}^2)_{0^{+},2^{+},4^{+}}$ with 50\%, 55.9\% and 67.9\% probabilities, respectively. The $6^{+}_1$ state has configuration $\pi(p_{1/2}^2g_{9/2}^8)\otimes\nu(d_{5/2}^1g_{7/2}^1)_{6^+}$. From these configurations, it is clear that the yrast $0^{+}_1-2^{+}_1-4^{+}_1-6^{+}_1$ states show neutron nature, which supports the results found in previous works \cite{Gorska1, Alber}. The little mixing is mainly due to the particle-hole character of the proton-neutron interaction. The neutron occupancies of the $2^{+}_{1}$ state are 1.259, 0.138, 0.446, 0.141, and 0.016 in the $d_{5/2}$, $s_{1/2}$, $g_{7/2}$, $d_{3/2}$ and $h_{11/2}$ orbitals, respectively. Here, the neutron $d_{5/2}$ orbital is dominating. We have also calculated the branching ratios for $8^{+}_1\rightarrow6^{+}_1$ and $8^{+}_1\rightarrow6^{+}_2$ transitions ($e_{\pi}=1.7e$, $e_{\nu}=1.1e$) as 99.96\% and 0.04\%, respectively. From these results, it seems that the $6^{+}_2$ state is fed by week $\gamma$-branches, quite similar to the calculated results in previous work \cite{Clark}.

	{\bf $^{102}$Cd}: The study of $^{102}$Cd (with $N=54$) helps for the understanding of $\pi\nu$ interaction in $\pi(g_{9/2})$, $\nu(d_{5/2})$, and $\nu (g_{7/2})$ orbitals \cite{Alber}. In $^{102}$Cd, the location of $2^{+}_1$ and $4^{+}_{1}$ states are comparable with the shell model. Experimentally, the half-life of $4^{+}_1$ state is less than 5.6 ps \cite{NNDC}. Our calculated half-life is 1.44 ps. This calculated half-life may be useful to compare experimental half-life once it is measured more accurately. In the Cd isotopic chain, the $^{102}$Cd isotope lies in the transitional region between single-particle character (as in nuclei $^{98}$Cd) and collective vibrational isotopes (with $N\ge56$). The $E2$ isomers with spin-parity $8^{+}$ are observed in this transitional region. Our calculated isomeric state $8^{+}_{1}$ is comparable with the experimental data. 
    The shell-model result shows that the $5^+_1$ and $6^+_1$ states locate close to each other. Since the experimental data show a $6^+$ state in the same energy region, the tentative $(5^+,6^+)$ state would be assigned  $5^+$.

	{\bf $^{104}$Cd}: The  $^{104}$Cd shows nearly spherical shape at low spins. For this isotope, the excited $2^{+}_{1}$, $4^{+}_{1}$ and $4^{+}_{2}$ states reproduced well by the shell model. From the experimental data, the half-life of $4^{+}_1$ state is less than 4.2 ps \cite{NNDC}. With the effective charges (1.7, 1.1)$e$, the calculated half-life is 2.37 ps. As reported in Ref. \cite{Angelis}, the $8^{+}_{1}$ state decays via $\gamma$-transition into $6^{+}_{1,2}$ states, this $8^{+}_{1}$ state is slightly compressed with our calculation. In the formation of $2^{+}_{1}$ state, the neutron occupancy is maximum in $d_{5/2}$ orbital with our calculation. We notice that no low-lying (approximately up to 3.5 MeV) states with negative parity are observed experimentally from $^{98}$Cd to $^{104}$Cd. It shows that there is no significant role of $h_{11/2}$ orbital in the filling of valence neutrons.

	{\bf $^{106}$Cd}: In the cadmium chain, $^{106}$Cd is a good example of the onset of collectivity \cite{Gray} with a simple structure. This isotope keeps a small but active number of valence nucleons. For this isotope, $2_1^{+}$, $4_1^{+}$, and $2^{+}_{2}$ states are reproduced well by our calculation. Earlier, Benczer-Koller $et$ $al.$ \cite{Koller} performed shell-model calculation for $^{106}$Cd considering $^{78}$Ni as a core. They have calculated the location of $2^+_1$ and $4^{+}_1$ states taking two sets of neutron model spaces. In our work, the calculated energy levels for $2^+_1$ and $4^+_1$ states are close to the experimental data than their results. The calculated wave function of the $2^{+}_{1}$ state is $\pi(p_{1/2}^2g_{9/2}^8)\otimes\nu(d_{5/2}^4g_{7/2}^4)$ with maximum probability 26.53\%. The configuration of octupole collective $3^{-}_1$ state is $\pi(p_{1/2}^2g_{9/2}^8)\otimes\nu(d_{5/2}^3g_{7/2}^4h_{11/2}^1)$. It is evident that there is a collective structure in $^{106}$Cd with the occupation of at least a single valence neutron in $h_{11/2}$ orbital.

	{\bf $^{108}$Cd}: For $^{108}$Cd, the $E(2^{+}_{1})$ state shows good agreement with the shell model calculation. Gade $et$ $al.$ \cite{Gade} established the intruder band in $^{108}$Cd isotope using the Doppler shift attenuation method. Our calculated $0^{+}_2$ intruder state lies 0.229 MeV lower than the experimental data. In $^{108}$Cd, the excited $4^{+}_1$ and $2^{+}_2$ states are two-phonon states. The $4^{+}_1$ state is reproduced well, but the $2^{+}_2$ state lies 0.154 MeV lower than the experimental level. The shell model calculated $3^{-}_{1}$ state shows good agreement with the experimental data. It is important to study the lowest quadrupole and octupole modes in $^{108}$Cd. The quadrupole-octupole coupled $(2^{+}\otimes3^{-})$ quintuplet state $5^{-}_1$ is underpredicted with our calculation. Another member of this family $1^{-}$ state lies 0.249 MeV higher than the experimental level.
 With our calculation, $0^{+}_{1}-2^{+}_{1}-4^{+}_{1}-2^{+}_2-0^{+}_2-3^{+}_1$ states are in the same order as in the experimental data.

	{\bf $^{110}$Cd}: In the case of $^{110}$Cd, the location of the calculated $2^{+}_1$ state is comparable with the experimental data. The $^{110}$Cd isotope is known as collective nucleus showing multiphonon excitations \cite{Nomura}. The order of 2-phonon states $2^{+}_2-4^{+}_1-0^{+}_3$ is reproduced with our calculation, but these states are compressed. The $3^{-}_1$ state is reproduced well with our calculation.

	{\bf $^{112}$Cd}: For $^{112}$Cd, the order of excited $0^{+}_1-2^{+}_1-4^{+}_1$ states are reproduced well with the shell model as in the experimental data. The first excited $2^{+}_1$ state lies  0.203 MeV lower than the experimental data with our calculation. Other excited states are compressed with our calculation. The energy level of the quadrupole-octupole coupled state $1^-$ should be near the sum of the energies of the quadrupole and octupole phonon states ($2^+_1$ and $3^-_1$, respectively) \cite{Garrett1}. The calculated $1^-_1$ state is underpredicted. But, the sum of the $2^+_1$ and $3^-_1$ energies (2.170 MeV) is close to the energy level of the $1^-_1$ state (2.108 MeV). It is expected that the negative parity states $5^{-}_1$ and $6^-_1$ are formed by the contribution of $h_{11/2}$ neutron orbital. From our calculation, the configuration of $5^{-}_1$ and $6^-_1$ states is $\pi(p_{1/2}^2g_{9/2}^8)\otimes\nu(d_{5/2}^4g_{7/2}^6d_{3/2}^1h_{11/2}^3)$. Both negative-parity states ($5^{-}_1$ and $6^-_1$) are coming from the coupling of valence neutrons in $h_{11/2}$ and $d_{3/2}$ orbitals.

	{\bf $^{114}$Cd}:  In the Cd chain, the $^{114}$Cd is situated in the middle of the neutron number $N=50$ and $N=82$. For this isotope, the excited energy levels are compressed with our calculation, but the order of $0^{+}_1-2^{+}_1-4^{+}_1-3^{+}_1$ states are reproduced well. It is expected that the $h_{11/2}$ neutron orbital plays a dominating role in the mid-shell isotopes of the Cd chain, even at low spin \cite{Jungclaus}. In the configuration of low-spin states (like $2^{+}_1$ state) of $^{112-114}$Cd, the $\nu(h_{11/2})$ orbital starts to dominate in our calculation.

	{\bf $^{116}$Cd}:  For  $^{116}$Cd, the order of $0^{+}_1-2^{+}_1-4^{+}_1-4^{+}_2$ states are reproduced with our calculation. But, these states are compressed. We can notice that the energy spectra are compressed with our calculation, but the location of $1^{+}_1$ state ($\pi(p_{1/2}^2g_{9/2}^8)\otimes\nu(d_{5/2}^6s_{1/2}^1g_{7/2}^6d_{3/2}^1h_{11/2}^4)$) is higher in comparison to other states.

	{\bf $^{118}$Cd}: In $^{118}$Cd, the calculated $2^{+}_1$ state lies 0.279 MeV lower than the experimental data. The location of 2-phonon states $4^+_1-2^+_2-0^+_2$ are also underpredicted with our calculation. The calculated $7^{-}_1$ state is comparable with the tentative experimental level at 1.269 MeV.
	The quadrupole interaction between protons and neutrons is responsible for the excitation energies of intruder states. Earlier, Wang $et$ $al.$ \cite{Wang} studied the intruder states in $^{118}$Cd. Our calculated intruder state $0^{+}_3$ is compressed by 0.396 MeV than the experimental data. As reported in Ref. \cite{Wang}, the intruder state $2^{+}_3$ (at 1916 keV) is compressed with shell model calculation.

	{\bf $^{120}$Cd}: For $^{120}$Cd, the order of $0^{+}_1-2^{+}_1-4^{+}_1$ states are reproduced well with our calculation as in the experimental data. But, the excited $2^{+}_1$ and $4^{+}_1$ states are compressed. The 2-phonon state $2^{+}_2$ lies 0.235 MeV lower than the experimental data with our calculation. We propose the configuration of this state as $\pi(p_{1/2}^2g_{9/2}^8)\otimes \nu (d_{5/2}^6s_{1/2}^2g_{7/2}^6d_{3/2}^2h_{11/2}^6)$ with 28.7\% probability, which shows the $2_{2}^+$ state comes from one proton pair breaking in $g_{9/2}$ orbital.

	{\bf $^{122}$Cd}: The low-lying states in $^{122}$Cd have been studied up to $J^{\pi}=8^{+}$ in Ref. \cite{Zamfir}. With our calculation, $3^{-}_{1}$ state is compressed by 0.268 MeV. But, in comparison to other states, the $3^-_1$ state ($\pi(p_{1/2}^2g_{9/2}^8)\otimes(d_{5/2}^6g_{7/2}^6d_{3/2}^3h_{11/2}^9)$) lies at high energy because it comes from the coupling of higher-lying neutron orbitals $\nu h_{11/2}$ and $\nu d_{3/2}$.

	{\bf $^{124}$Cd}: The level scheme of $^{124}$Cd has been reported in previous works \cite{Stoyer, Vancraeyenest}. The $5^{-}_{1}$ state is compressed by 0.259 MeV with our calculation. The order of excited $0^{+}_{1}$-$2^{+}_{1}$-$4^{+}_{1}$-$2^{+}_{2}$ states are well reproduced by the shell model calculation. Like $^{120,122}$Cd, the $h_{11/2}$ orbital is dominant in the formation of $2^{+}_{1}$ state.

	{\bf $^{126}$Cd}: Kautzsch $et$ $al.$ \cite{Kautzsch}  calculated the excited energy levels for $^{126}$Cd using shell-model. In our study, the calculated $2^{+}_{1}$ state is compressed by 0.270 MeV with the experimental data for this isotope. The $5^{-}_{1}$ state is comparable with our calculation. The proposed configuration of $5^{-}_1$ state is $\pi(p_{1/2}^2g_{9/2}^8)\otimes\nu(d_{5/2}^6s_{1/2}^1g_{7/2}^8d_{3/2}^2h_{11/2}^{11})$ with 28.6\% probability. Here, the $5^{-}_1$ state arises from the coupling of unpaired neutrons in $s_{1/2}$ and $h_{11/2}$ orbitals. The energy spectra of all states are in the same order with our calculation. Hoteling $et$ $al.$ \cite{Hoteling} suggested a new energy state $7^-$ at 1950 keV. With our calculation, this state is very close to the tentative experimental level with an energy difference of 15 keV. This state comes from  $\pi(p_{1/2}^2g_{9/2}^8)\otimes\nu(d_{5/2}^6s_{1/2}^2g_{7/2}^8d_{3/2}^3h_{11/2}^{9})$ configuration with 59.3\% probability.

	{\bf $^{128}$Cd}: For $^{128}$Cd, the shell model predicted $2^{+}_1$ state is comparable with the experimental data. From the dominant configuration of this state ($\pi(p_{1/2}^2g_{9/2}^8)\otimes\nu(d_{5/2}^6s_{1/2}^2g_{7/2}^8d_{3/2}^4h_{11/2}^{10})$), it is clear that this state comes with one proton pair breaking in $g_{9/2}$ orbital. The $4^{+}_{1}$ state lies 0.278 MeV lower than the experimental data with our calculation. The order of $8^{+}_{1}-10^{+}_{1}$ states is reversed with the calculated spectra. But, the energy gap between $8^{+}_{1}$ and $10^{+}_{1}$ is very close as in the experimental data. As expected, the positive-parity states $0^{+}_1-8^+_1$ are showing predominent $\pi(g_{9/2}^{-2})$ character from our calculation. The two negative parity states $5^-_1$ and $7^-_1$ are coming from $\pi(p_{1/2}^2g_{9/2}^8)\otimes\nu(d_{5/2}^6s_{1/2}^2g_{7/2}^8d_{3/2}^3h_{11/2}^{11})$ configuration with 39.3\% and 84.9\% probabilities, respectively. Both states are showing $\nu(d_{3/2}^{-1}h_{11/2}^{-1})$ character.
	
	\begin{figure}[h]
		\includegraphics[width=80mm]{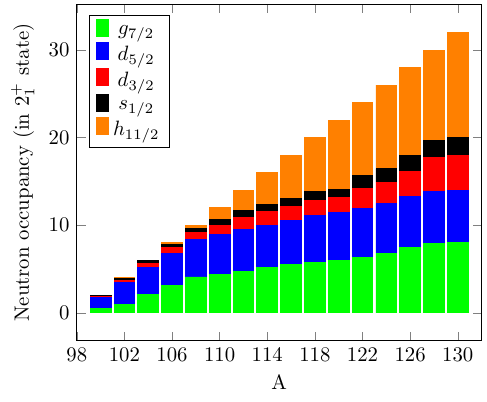}
		\caption{\label{Occupancy} The occupancy for different neutron orbitals corresponding to excited $2^{+}_1$ state in the Cd isotopic chain with the G-matrix interaction.}
	\end{figure}

	{\bf $^{130}$Cd}: In $^{130}$Cd, the order of calculated energy spectra is the same as in the experimental data. The calculated ground state $0^+_1$ shows $\pi (g_{9/2}^{-2})$ nature. The location of $2^{+}_{1}$ and $4^{+}_{1}$ states reproduced very well by the shell model. The $6^{+}_{1}$ and $8^{+}_{1}$ states lie slightly higher with our calculation. From our calculations, the $4^+_1$, $6^+_1$ and $8^+_1$ states are showing $\pi (g_{9/2}^{-2})$ character. We have also calculated excited $5^{-}_1$ and $4^{-}_1$ states, which are not observed experimentally yet. Our results for these two states may help to compare new experimental data in the future.
	
	We can notice that our calculated energy spectra are deviating from the experimental data from $^{110}$Cd and start to compress up to $^{120}$Cd. The reason behind this is that when we perform shell-model calculations using realistic interactions and increase the number of particles (neutrons), our results deteriorate rapidly \cite{Caurier}.  The discrepancies  might be due to monopole part of the effective interaction and missing degrees of freedom in the valence space.

	\begin{figure}
     \includegraphics[width=80mm]{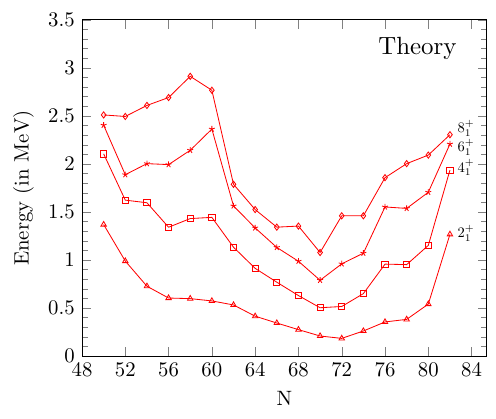}
     \includegraphics[width=80mm]{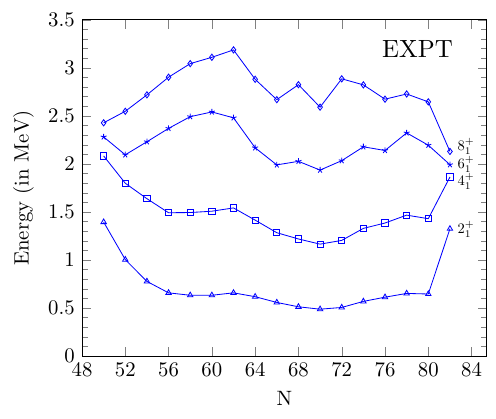}
		\caption{ Comparison between calculated and the experimental excitation energy of the $2^{+}_{1}$, $4^{+}_{1}$, $6^{+}_{1}$ and $8^{+}_{1}$ states for even $^{98-130}$Cd isotopes.}
		\label{figE2}
	\end{figure}
	
	In Fig. \ref{Occupancy}, we have shown the occupancy of different neutron orbitals corresponding to $2^{+}_1$ state from $N=50$ to $N=82$ in the Cd isotopic chain. The $d_{5/2}$ and $g_{7/2}$ orbitals are dominant in the neutron-deficient region. With the increase of neutron numbers, occupancy increases in all orbitals. In the mid of the Cd chain, the dominance of the $h_{11/2}$ orbital can be observed. As we further increase the neutron number, the occupancy of $h_{11/2}$ orbital becomes more than the other orbitals, which shows the importance of $h_{11/2}$ orbital in the formation of $2^{+}_1$ state in neutron-rich Cd isotopes.
	
	\subsubsection{\bf Variation in the excitation energy of $2^{+}_{1}$, $4^{+}_{1}$, $6^{+}_{1}$, and $8^+_1$ states from $N=50$ to $N=82$}
	
	Here, we have shown the variation of excitation energy of the $2^{+}_1$, $4^{+}_{1}$, $6^{+}_{1}$, and $8^+_1$ states for the Cd isotopic chain. The trend for excited $2^{+}_{1}$ energy states for $N=50-82$ reproduced well with the shell model calculation. We can notice that there is a small change at $N=62$ in the variation of the experimental excitation energy of $2^{+}_{1}$ state, as shown in Fig. \ref{figE2}. The cause may be a sub-shell gap in the single-particle energies of neutron orbitals. When neutrons start occupying in the valence space, first $d_{5/2}$, $g_{7/2}$ and $s_{1/2}$ orbitals prefer to filling. As we further increase the neutrons around $N=62$, the dominance of $h_{11/2}$ orbital will be started. So, the change in variation of $2^{+}_{1}$ state energy is related to the change in the filling of neutron orbitals. At $N=82$, there is the neutron shell closure. Hence, the energy of the excited $2^{+}_{1}$ state suddenly increases at $N=82$.
	
	A similar trend is also following in the variation of $4^+_1$ state energy like $2^+_1$ state. However, the $4^+_1$ states are compressed in the mid-shell region. In the case of $6^+_1$ and $8^+_1$ states, our results are good for neutron-deficient isotopes and the shell closure at $N=82$, but these states are compressed in the mid-shell like $4^+_1$ states.
	
	\begin{figure}[h]
		\includegraphics[width=80mm]{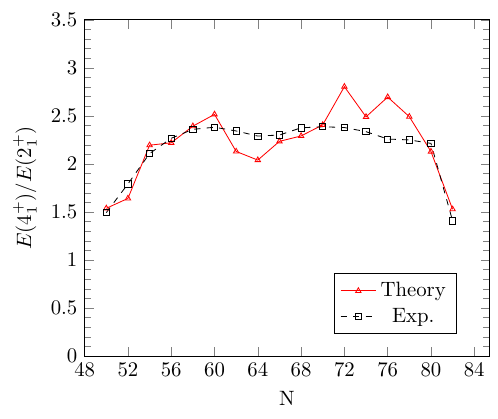}
		\caption{ Comparison of the  theoretical and experimental excitation energy ratios of $4^{+}_1$ and $2^{+}_1$ states in Cd chain. The experimental data are taken from \cite{NNDC}.}
		\label{E42}
	\end{figure}
	
	\subsubsection{\bf The $4^{+}_1/2^{+}_1$ energy ratio ($R_{4/2}$)}
	
	In Fig. \ref{E42}, we have compared the experimental and theoretical excitation energy ratios of $4^{+}_1$ and $2^{+}_1$ states in the Cd isotopic chain. We can notice that the $4^{+}_1/2^{+}_1$ energy ratio ($R_{4/2}$) follows the same trend from the $N=50$ closed shell toward more collective structures with the shell model calculation as in the experimental data. The value of $R_{4/2}$ ratio is less than $2.0$ for $^{98,100}$Cd isotopes with our calculation, which corresponds to a single-particle structure near closed shells. For $^{102-108}$Cd isotopes, the calculated $R_{4/2}$ ratio lies between 2.0 and 2.6, which shows that these isotopes are quasi-deformed. For $^{106}$Cd, the experimental value of $R_{4/2}$ ratio is 2.360. From our calculation, the $R_{4/2}$ ratio is 2.393. The predicted $R_{4/2}$ ratio from the pure vibrational and rotational model are 2.00 and 3.33, respectively \cite{Koller}. Here, our calculated result shows good agreement with the experimental data. For $^{108}$Cd, this ratio is 2.516. But, there is a decrement in the $R_{4/2}$ ratio at $N=62$, which indicates $^{108}$Cd has large deformation. However, our results are staggering from $N=70$ to $N=78$.
  In $^{120,124}$Cd, the main configurations of $4^{+}_1$ states are $\pi(g_{9/2}^{-2})\otimes\nu(d_{5/2}^6g_{7/2}^6d_{3/2}^2h_{11/2}^8)$ ($\sim 25.1$\%) and $\pi(g_{9/2}^{-2})\otimes\nu(d_{5/2}^6s_{1/2}^2g_{7/2}^6d_{3/2}^2h_{11/2}^{10})$ ($\sim 22.7$\%) with our calculation, respectively. In both configurations, there are no unpaired nucleons in valence proton and neutron orbitals. While, the configuration of $4^+_1$ states in neighbouring isotopes $^{118}$Cd [$\pi(g_{9/2}^{-2})\otimes\nu({d_{5/2}^6s_{1/2}^1g_{7/2}^6d_{3/2}^1}h_{11/2}^6)$ ($\sim 17.4$\%)] and $^{122}$Cd [$\pi(g_{9/2}^{-2})\otimes\nu({d_{5/2}^6s_{1/2}^1g_{7/2}^6d_{3/2}^3}h_{11/2}^8)$ ($\sim 22.2$\%)] have unpaired neutrons. So, it needs more energy to form the $4^{+}_1$ states in $^{120,124}$Cd in comparison to $^{118,122}$Cd, which may be a possible reason for the sudden jump in the $4^+_1$ states and increased $R_{4/2}$ ratio for $^{120,124}$Cd isotopes.

	\subsection{Electromagnetic properties}
	
	\begin{figure}
		\includegraphics[width=81mm]{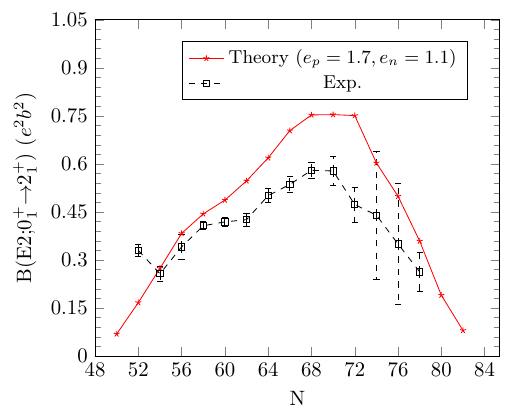}
		\caption{ Variation of $B(E2;0^+_1\rightarrow2^+_1)$ values along Cd isotopic chain. The experimental data are taken from \cite{Pritychenko, Maheshwari}.}
		\label{BE2fig}
	\end{figure}
	
	\begin{figure}
		\includegraphics[width=81mm]{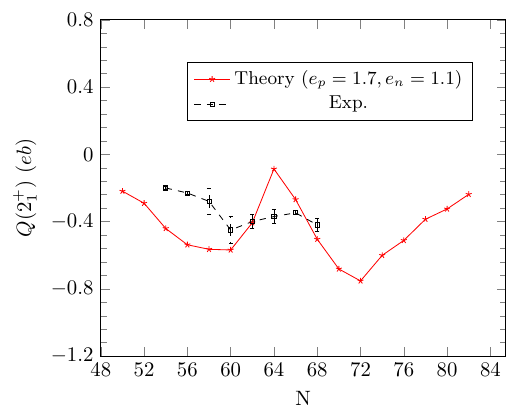}
		\caption{ Electric quadrupole moments for $2^{+}_{1}$ state in Cd isotopic chain. The experimental data for $^{102,104}$Cd and $^{106-116}$Cd are taken from \cite{Ekstrom} and \cite{NDS}, respectively.}
		\label{electric}
	\end{figure}

	\begin{table*}
		\caption{Comparison of theoretical and experimental $B(E2)$ transition probabilities in cadmium isotopes.  The units are in $e^2$b$^2$. In the present work, we have taken effective charges ($e_p$, $e_n$)$=$(1.7, 1.1)$e$ from \cite{Boelaert}.}
		\begin{tabular}{cccc}
			\hline
			Isotope	&	$J_i^{\pi} \rightarrow J_f^{\pi}$  & G-matrix  &	Exp. 	\T\B \\
			\hline
			$^{98}$Cd	&	$0^{+}_{1}$$\rightarrow$$2^{+}_{1}$               &  0.069  &  -  \T\B \\
			&   $2^{+}_{1}$$\rightarrow$$4^{+}_{1}$             &  0.033     &  - \T\B \\ 
			&   $4^{+}_{1}$$\rightarrow$$6^{+}_{1}$             &  0.018  &  0.018(3)\T\B \\
			$^{100}$Cd	&	$0_{1}^{+}$$\rightarrow$$2_1^{+}$   &  0.167  &  0.33(2) \T\B \\
			&   $2_{1}^{+}$$\rightarrow$$4_{1}^{+}$       &  0.047  &  - \T\B \\
			&   $4_{1}^{+}$$\rightarrow$$6_{1}^{+}$   &  0.021  &  - \T\B \\ 
			$^{102}$Cd	&	$0_{1}^{+}$$\rightarrow$$2_1^{+}$     &  0.274  &  0.257(23)  \T\B \\
			&   $2_{1}^{+}$$\rightarrow$$4_{1}^{+}$     &  0.143  &  $>$0.038 \T\B \\
			&   $4_{1}^{+}$$\rightarrow$$6_{1}^{+}$  &  0.001  &  0.057(45)  \T\B \\
			$^{104}$Cd	&	$0_{1}^{+}$$\rightarrow$$2_1^{+}$      & 0.383 &  0.341(40)  \T\B \\
			&   $2_{1}^{+}$$\rightarrow$$4_{1}^{+}$    & 0.198 & 0.058 $>$ \T\B \\
			&   $4_{1}^{+}$$\rightarrow$$6_{1}^{+}$    & 0.154  & 0.037 $>$ \T\B \\	
			$^{106}$Cd	&	$0_{1}^{+}$$\rightarrow$$2_{1}^{+}$     & 0.444  &  0.407(12) \T\B \\
			&   $2_{1}^{+}$$\rightarrow$$4_{1}^{+}$   &  0.230  & 0.247(32) \T\B \\
			&   $4_{1}^{+}$$\rightarrow$$6_{1}^{+}$   &  0.021  & - \T\B \\
			$^{108}$Cd	&	$0_{1}^{+}$$\rightarrow$$2_{1}^{+}$     & 0.487  &  0.419(14) \T\B \\
			&   $2_{1}^{+}$$\rightarrow$$4_{1}^{+}$   & 0.255 & 0.225(33) \T\B \\
			$^{110}$Cd	&	$0_{1}^{+}$$\rightarrow$$2_{1}^{+}$   & 0.547  &   0.426(21) \T\B \\
			&   $2_{1}^{+}$$\rightarrow$$4_{1}^{+}$   & 0.228  &  0.237(51) \T\B \\
			$^{112}$Cd	&	$0_{1}^{+}$$\rightarrow$$2_{1}^{+}$    & 0.619  &  0.501(22) \T\B \\
			&   $2_{1}^{+}$$\rightarrow$$4_{1}^{+}$   & 0.303  & 0.364(46) \T\B \\
			$^{114}$Cd	&	$0_{1}^{+}$$\rightarrow$$2_{1}^{+}$    & 0.704  &  0.536(25) \T\B \\
			$^{116}$Cd	&	$0_{1}^{+}$$\rightarrow$$2_{1}^{+}$    & 0.753  &  0.580(26) \T\B \\
			$^{118}$Cd	&	$0_{1}^{+}$$\rightarrow$$2_{1}^{+}$    & 0.754  &  0.578(44) \T\B \\
			$^{120}$Cd	&	$0_{1}^{+}$$\rightarrow$$2_{1}^{+}$    & 0.751  &  0.473(55) \T\B \\
			$^{122}$Cd	&	$0_{1}^{+}$$\rightarrow$$2_{1}^{+}$    & 0.603  &  0.44(20)  \T\B \\
			$^{124}$Cd	&	$0_{1}^{+}$$\rightarrow$$2_{1}^{+}$    & 0.499  &  0.35(19)  \T\B \\
			$^{126}$Cd	&	$0_{1}^{+}$$\rightarrow$$2_{1}^{+}$    & 0.359  &  0.263(60) \T\B \\
			$^{128}$Cd	&	$0_{1}^{+}$$\rightarrow$$2_{1}^{+}$    & 0.190  &  - \T\B \\
			$^{130}$Cd	&	$0_{1}^{+}$$\rightarrow$$2_{1}^{+}$    & 0.080  &  - \T\B \\
			\hline
		\end{tabular}
		\label{BE2}
	\end{table*}
	
	Here, we have discussed reduced transition probabilities, quadrupole, and magnetic moments for even Cd isotopes. We have explained the behavior of $B(E2;0^+_1\rightarrow2^+_1)$ transition for Cd isotopes from $N=50$ to $N=82$. The $B(E2)$ values of the Cd and Sn isotopic chain show similar trends \cite{Siciliano}. In the experimental data of $B(E2)$ values of the Cd chain, there is a dip near $N=62$, as shown in Fig. \ref{BE2fig}. As shown in Fig. \ref{fig1}, the ESPEs for $\nu(d_{5/2})$, $\nu(g_{7/2})$ and $\nu(s_{1/2})$ orbitals lie lower than the $\nu(h_{11/2})$ orbital. The asymmetry near $N=62$ is because before $N=62$, $d_{5/2}$ orbital is dominating, and after that, $h_{11/2}$ orbital is dominating. Our results are in good agreement with the experimental trend. We have used effective charges $(e_{\pi}=1.7, e_{\nu}=1.1)$ from Ref. \cite{Boelaert} for calculating $B(E2; 0^{+}_1\rightarrow 2^{+}_1)$ values.
	
	\begin{figure}
		\includegraphics[width=81mm]{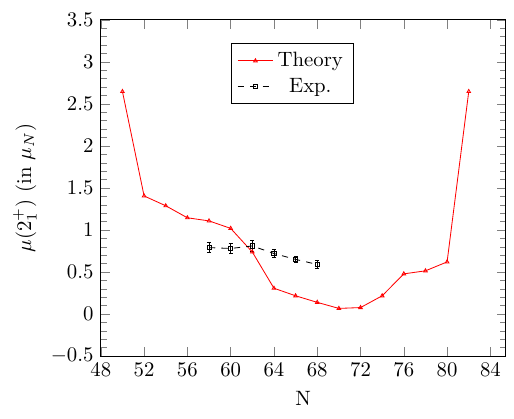}
		\caption{ Magnetic moments for $2^{+}_{1}$ state in Cd isotopic chain. The experimental data are taken from \cite{NDS}.}
		\label{magnetic}
	\end{figure}
	
	\begin{table}
		\centering
		\caption{Comparison of theoretical and experimental electric quadrapole  and magnetic moments  in cadmium isotopes. In our calculations we have taken effective charges ($e_p$, $e_n$)$=$(1.7, 1.1). The $g$-values are taken as $g_s^p=3.91$, and $g_s^n=-2.678$.}
		\begin{tabular}{lcccccc}
			\hline
			&                     &     & \multicolumn{2}{c}{Q (eb)}    & \multicolumn{2}{c}{~$\mu$ ($\mu_N$)} \T\B\\
			\cline{4-5}
			\cline{6-7}	
			
			Nuclei  & $A$  & $J^{\pi}$  & G-matrix &  Exp.  & G-matrix & Exp.\T\B\\
			\hline
			Cd      & 98  & $2^+_{1}$   & -0.219  & -   & 2.647  & -\T\B\\
			&  & $4^{+}_{1}$  & -0.139  & -  & 5.293 & -\T\B\\
			&  & $6^{+}_{1}$  & 0.088   & -  & 7.940 & -\T\B\\
			&  & $8^{+}_{1}$  & 0.438   & - & 10.587 & -\T\B\\
			
			& 100 &$2^+_{1}$  & -0.291  & - & 1.405 & -\T\B\\
			&  & $4^{+}_{1}$  & -0.361  & - & -0.232 & -\T\B\\
			&  & $6^{+}_{1}$  & -0.572  & - & 0.220 & -\T\B\\
			&  & $8^{+}_{1}$  & 0.659   & - & 10.154 & 9.9(5)\T\B\\
			
			& 102 &$2^+_{1}$  & -0.441  & -0.20(1)  & 1.288 & -\T\B\\
			&  & $4^{+}_{1}$  & -0.527  & -  & 2.436 & -\T\B\\
			&  & $6^{+}_{1}$  & -0.473  & -  & 0.572 & -\T\B\\
			&  & $8^{+}_{1}$  & 0.738   & 0.76(9) & 9.117 & 10.3(2)\T\B\\
			
			& 104 &$2^+_{1}$  & -0.538  & -0.23(1)  & 1.145 & -\T\B\\
			&  & $4^{+}_{1}$  & -0.675  & -  & 1.844 & -\T\B\\
			
			& 106 &$2^+_{1}$  & -0.565  & -0.28(8) & 1.108 & 0.79(6)\T\B\\
			&  & $4^{+}_{1}$  & -0.691  & - & 1.935 & 0.9(2)\T\B\\
			
			& 108 &$2^+_{1}$  & -0.570  & -0.45(8)  & 1.019 & 0.78(6)\T\B\\
			&  & $4^{+}_{1}$  & -0.604  & - & 1.781 & -\T\B\\
			
			& 110 &$2^+_{1}$  & -0.407  & -0.40(4) & 0.738 & 0.81(6)\T\B\\
			&  & $4^{+}_{1}$  & -0.217  & - & 0.295 & -\T\B\\
			
			& 112 &$2^+_{1}$  & -0.087  & -0.37(4) & 0.307 & 0.72(5)\T\B\\
			&  & $4^{+}_{1}$  & -0.157  & - & 0.179 & -\T\B\\
			
			& 114 &$2^+_{1}$  & -0.269  & -0.348(12)  & 0.217 & 0.65(4)\T\B\\

			& 116 &$2^+_{1}$  & -0.505  & -0.42(4)  & 0.139 & 0.59(5)\T\B\\

			& 118 &$2^+_{1}$  & -0.682  & -  & 0.067 & -\T\B\\

			& 120 &$2^+_{1}$  & -0.753  & - & 0.077 & -\T\B\\

			& 122 &$2^+_{1}$  & -0.601  & - & 0.217 & -\T\B\\

			& 124 &$2^+_{1}$  & -0.512  & -  & 0.478 & -\T\B\\

			& 126 &$2^+_{1}$  & -0.386  & -  & 0.514 & -\T\B\\

			& 128 &$2^+_{1}$  & -0.325  & - & 0.620 & -\T\B\\

			& 130 &$2^+_{1}$  & -0.238  & - & 2.647 & -\T\B\\
		
			\hline
		\end{tabular}
		\label{Moments}
	\end{table}
	
	\begin{table}
		\caption{Comparison between theoretical and experimental $B(M1)$ (in $\mu_N^{2}$) values in cadmium isotopes. In our calculations we have taken $g_s^p=3.91$, $g_s^n=-2.678$.}
		\begin{tabular}{cccc}
			\hline
			Isotope	&	$J_i^{\pi} \rightarrow J_f^{\pi}$  & G-matrix &	Exp.	\T\B \\
			\hline
			$^{98}$Cd	&	$2_{2}^{+}$$\rightarrow$$2_{1}^{+}$     &  0.818 &  - \T\B \\
			&   $4_{2}^{+}$$\rightarrow$$4_{1}^{+}$    &  0.477  &  - \T\B \\ 
			$^{100}$Cd &	$2_{2}^{+}$$\rightarrow$$2_{1}^{+}$      &  0.033 &  - \T\B \\
			&   $4_{2}^{+}$$\rightarrow$$4_{1}^{+}$    &  0.134 &  -  \T\B \\
			$^{102}$Cd	&	$2_{2}^{+}$$\rightarrow$$2_{1}^{+}$    &  0.043  &  - \T\B \\
			&   $5_{1}^{+}$$\rightarrow$$4_{1}^{+}$  &  0.002  &  - \T\B \\
			$^{104}$Cd  &   $2_{2}^{+}$$\rightarrow$$2_{1}^{+}$    &  0.043  &  -  \T\B \\
			&   $4_{2}^{+}$$\rightarrow$$4_{1}^{+}$  & 0.069 &  -  \T\B \\
			$^{106}$Cd  &   $2_{2}^{+}$$\rightarrow$$2_{1}^{+}$    &  0.015  &  0.013(3) \T\B \\
			&   $4_{2}^{+}$$\rightarrow$$4_{1}^{+}$   &  0.001  &  $>$0.047  \T\B \\
			$^{108}$Cd  &   $2_{2}^{+}$$\rightarrow$$2_{1}^{+}$  & 0.025 &  0.014(9) \T\B \\
			&   $6_{1}^{-}$$\rightarrow$$5_{1}^{-}$  & 0.021 & 0.0027(18)  \T\B \\
			
			$^{110}$Cd  &   $2_{2}^{+}$$\rightarrow$$2_{1}^{+}$   & 0.095 &  0.024(4) \T\B \\
			&   $4_{3}^{+}$$\rightarrow$$4_{1}^{+}$   &  0.002  &  0.104(+30-45)  \T\B \\
			$^{112}$Cd  &   $3_{1}^{+}$$\rightarrow$$2_{1}^{+}$  &  0.0004  &  0.003(1)  \T\B \\
			$^{116}$Cd  &   $4_{2}^{+}$$\rightarrow$$4_{1}^{+}$   &  0.001  & 0.11(+7-11)  \T\B \\
			$^{118}$Cd  &   $2_{2}^{+}$$\rightarrow$$2_{1}^{+}$   & 0.010 & - \T\B \\
			$^{120}$Cd  &   $2_{2}^{+}$$\rightarrow$$2_{1}^{+}$   & 0.0004 & - \T\B \\
			$^{122}$Cd  &   $2_{2}^{+}$$\rightarrow$$2_{1}^{+}$   &  0.003  &  -  \T\B \\
			$^{124}$Cd  &   $1_{1}^{+}$$\rightarrow$$0_{1}^{+}$   &  0.0002  &  -  \T\B \\
			$^{126}$Cd  &   $2_{2}^{+}$$\rightarrow$$2_{1}^{+}$   &  0.0001  &  -  \T\B \\
			\hline
		\end{tabular}
		
		\label{B(M1)}
	\end{table}

	We have also calculated $B(E2)$ transitions for other states, reported in Table \ref{BE2}. In $^{98}$Cd, $B(E2;4^+_1\rightarrow6^+_1)$ value is reproduced well and equals to 0.018 $e^2b^2$ with our calculation. Park $et$ $al.$ \cite{Park} studied the $6^{+}_1$ isomeric state in a decay spectroscopy experiment with half-life, $T_{1/2}=13(2)$ ns. With our calculation, the calculated half-life of the $6^{+}_1$ state is 1.92 ns. In the case of even $^{106-112}$Cd isotopes, the calculated $B(E2;2^{+}_1 \rightarrow 4^{+}_1)$ values are showing good agreement with the experimental data \cite{NNDC}.

	The trend of the $11/2^-$ quadrupole moments for odd $^{111-129}$Cd isotopes has been studied in previous works \cite{Yordanov, Yang}. In $^{106-116}$Cd isotopes, the experimental data are available for $Q(2^{+}_1)$ values. We have used shell-model interpreted data of $Q(2^{+}_1)$ \cite{Ekstrom} for $^{102,104}$Cd. We have compared our results with the experimental data for even Cd isotopic chain as shown in Fig. \ref{electric}. One can notice that the $Q(2^{+}_1)$ values show nearly constant behavior for $^{102-106}$Cd isotopes. Experimentally, there is a dip of $Q(2^{+}_1)$ value at $^{108}$Cd. After that, the trend of $Q(2^{+}_1)$ is almost constant up to $^{116}$Cd.  For $^{98,100}$Cd isotopes, $Q(2^{+}_1)$ values are close to each other with our calculation. But, they have not been observed in the experiment yet. Experimental data are not available for the electric quadrupole moment of $2^{+}_1$ state in neutron-rich even Cd isotopes. In the future, our results may be helpful in comparing upcoming experimental results in Cd isotopes.
	
	For $^{102}$Cd, the experimental $Q(8^{+}_1)$ value is close to our result. We have also calculated electric quadrupole moments for other states, as reported in Table \ref{Moments}.
	
	For magnetic moments of $2^{+}_1$ state, the experimental data are available only for $^{106-116}$Cd isotopes like quadrupole moments. The trend of our calculated results for $\mu(2^{+}_1)$ values is not similar to the experimental data as shown in Fig. \ref{magnetic}. But, some calculated $\mu(2^{+}_1)$ values are close to the experimental data. In the case of $^{100,102}$Cd isotopes, the experimental $\mu(8^{+}_1)$ show good agreement with our results. For $^{106}$Cd, the $\mu(4^{+}_1)$ value is two times higher than the experimental value with our result. In Table \ref{Moments}, we have also reported magnetic moments for other states where the experimental data are not available.
	
	For the even Cd isotopic chain, the experimental data of $B(M1)$ values are available only for a few states. We have reported $B(M1)$ transitions for these isotopes in Table \ref{B(M1)}. 
	Overall, the $B(M1;2^{+}_2\rightarrow 2^{+}_1)$ transitions are well described, but the other transitions not so much.
	
		\begin{figure}
			\includegraphics[width=81mm]{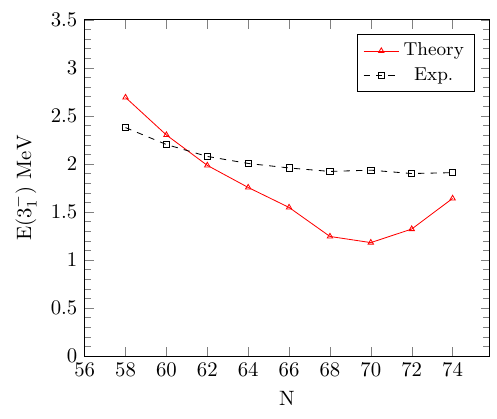}
			\caption{\label{E3-} Comparison of the experimental excitation energy of the octupole collective $3^{-}_{1}$ states for $^{106-122}$Cd isotopes with our results.}
		\end{figure}
		
		\begin{table}
			\caption{\label{table5} The calculated shell model wave functions corresponding to $3^{-}_{1}$ state in even-even $^{106-122}$Cd isotopes.}
			\begin{tabular}{cccc}
				\hline \hline
				Nuclei     &  ~~ $J^{\pi}$   & ~~ Probability  & Configuration \\
				\hline \hline \\
				$^{106}$Cd & ~~ $3^-_1$  & 31.8\%  &  $\pi(p_{1/2}^2g_{9/2}^8)\otimes\nu(d_{5/2}^4g_{7/2}^3h_{11/2}^1)$ \\
				&    & 10.0\%  &  $\pi(p_{1/2}^2g_{9/2}^8)\otimes\nu(d_{5/2}^3g_{7/2}^4h_{11/2}^1)$ \\
				\hline \\
				$^{108}$Cd &    & 15.8\%  &  $\pi(p_{1/2}^2g_{9/2}^8)\otimes\nu(d_{5/2}^4g_{7/2}^5h_{11/2}^1)$ \\
				&    &  7.7\%  &  $\pi(p_{1/2}^2g_{9/2}^8)\otimes\nu(d_{5/2}^4g_{7/2}^4d_{3/2}^1h_{11/2}^1)$ \\
				&    &  7.4\%  &  $\pi(p_{1/2}^2g_{9/2}^8)\otimes\nu(d_{5/2}^4g_{7/2}^3d_{3/2}^2h_{11/2}^1)$ \\
				
				\hline \\
				$^{110}$Cd &    & 8.3\%   & $\pi(p_{1/2}^2g_{9/2}^8)\otimes\nu(d_{5/2}^6s_{1/2}^1g_{7/2}^4h_{11/2}^1)$ \\
				&    & 6.9\%   & $\pi(p_{1/2}^2g_{9/2}^8)\otimes\nu(d_{5/2}^4s_{1/2}^1g_{7/2}^6h_{11/2}^1)$ \\
				&    & 6.6\%   & $\pi(p_{1/2}^2g_{9/2}^8)\otimes\nu(d_{5/2}^5g_{7/2}^5d_{3/2}^1h_{11/2}^1)$ \\
				
				\hline \\
				$^{112}$Cd &    & 15.7\%   &  $\pi(p_{1/2}^2g_{9/2}^8)\otimes\nu(d_{5/2}^4g_{7/2}^6d_{3/2}^1h_{11/2}^3)$ \\
				&    & 9.1\%   &  $\pi(p_{1/2}^2g_{9/2}^8)\otimes\nu(d_{5/2}^5g_{7/2}^5d_{3/2}^1h_{11/2}^3)$ \\
				&    & 7.6\%   &  $\pi(p_{1/2}^2g_{9/2}^8)\otimes\nu(d_{5/2}^6g_{7/2}^4d_{3/2}^1h_{11/2}^3)$ \\
				
				\hline \\
				$^{114}$Cd &    & 13.2\%   &  $\pi(p_{1/2}^2g_{9/2}^8)\otimes\nu(d_{5/2}^5g_{7/2}^6d_{3/2}^2h_{11/2}^3)$ \\
				&    & 9.8\%   &  $\pi(p_{1/2}^2g_{9/2}^8)\otimes\nu(d_{5/2}^6g_{7/2}^5d_{3/2}^2h_{11/2}^3)$ \\
				&    & 8.4\%   &  $\pi(p_{1/2}^2g_{9/2}^8)\otimes\nu(d_{5/2}^5s_{1/2}^1g_{7/2}^6d_{3/2}^1h_{11/2}^3)$ \\
				
				\hline \\
				$^{116}$Cd &    & 23.2\%  & $\pi(p_{1/2}^2g_{9/2}^8)\otimes\nu(d_{5/2}^5g_{7/2}^6d_{3/2}^2h_{11/2}^5)$ \\
				&    & 15.7\%  & $\pi(p_{1/2}^2g_{9/2}^8)\otimes\nu(d_{5/2}^5s_{1/2}^1g_{7/2}^6d_{3/2}^1h_{11/2}^5)$ \\
				
				\hline \\
				$^{118}$Cd &    & 27.7\%  & $\pi(p_{1/2}^2g_{9/2}^8)\otimes\nu(d_{5/2}^5g_{7/2}^6d_{3/2}^2h_{11/2}^7)$ \\
				&    & 18.3\%  & $\pi(p_{1/2}^2g_{9/2}^8)\otimes\nu(d_{5/2}^5s_{1/2}^1g_{7/2}^6d_{3/2}^1h_{11/2}^7)$ \\
				
				\hline \\
				$^{120}$Cd &    & 37.3\%  &  $\pi(p_{1/2}^2g_{9/2}^8)\otimes\nu(d_{5/2}^6s_{1/2}^1g_{7/2}^6d_{3/2}^2h_{11/2}^7)$ \\
				&    & 24.5\%  &  $\pi(p_{1/2}^2g_{9/2}^8)\otimes\nu(d_{5/2}^6s_{1/2}^2g_{7/2}^6d_{3/2}^1h_{11/2}^7)$ \\
				\hline \\
				$^{122}$Cd &    & 27.3\%  & $\pi(p_{1/2}^2g_{9/2}^8)\otimes\nu(d_{5/2}^6g_{7/2}^6d_{3/2}^3h_{11/2}^9)$ \\
				&    & 22.9\%  & $\pi(p_{1/2}^2g_{9/2}^8)\otimes\nu(d_{5/2}^6s_{1/2}^1g_{7/2}^6d_{3/2}^2h_{11/2}^9)$ \\
				\hline \hline
				
			\end{tabular}
		\end{table}
		
		\subsection{Octupole collective $3^{-}$ states in even-even Cd isotopes}
		
		The study of $3^{-}$ states is important for the understanding of the octupole vibrational mode of a nucleus. Low-lying $3^{-}$ octupole states are found in $^{106-122}$Cd isotopes. They decay via electromagnetic transitions to $0^+$ and $2^+$ states. In Fig. \ref{E3-}, we have compared the experimental trend of the $3^-$ state in the Cd chain with our results.
		
		As shown in Fig. \ref{E3-}, these $3^-_1$ states are showing reasonable agreement with the experimental data in the $^{106-112}$Cd isotopes. As we move in the neutron-rich region, their locations are compressed up to $^{120}$Cd.  The major contributions of this state are coming with $\nu (d_{5/2}^{-1} h_{11/2}^{1})$ configuration for $^{114-118}$Cd isotopes. For $^{122}$Cd, the result is slightly better.
  Further, the inclusion of proton orbital $p_{3/2}$ may improve the trend for energy. From Table \ref{table5}, it is also clear that the configurations of $3^-_1$ state are fragmented in $^{108-114}$Cd, and the 
  largest probabilities of the configuration in the wavefunction are less than 25\% for these isotopes. It means that the nucleon-nucleon interaction causes a large configuration mixing.
		
        \begin{table}[h]
			\centering
			\caption{Comparison of theoretical and experimental $B(E2)$ transitions (in $e^2b^2$) of $8^+_1$ isomeric states in Cd isotopes. The calculated partial half-lives are also reported. The effective charges are taken as ($e_p$, $e_n$)$=$(1.7, 1.1)e.}
			\begin{tabular}{lccccc}
				\hline
				&                 & \multicolumn{2}{c}{$B(E2)$}    & \multicolumn{2}{c}{$T_{1/2}$} \T\B\\
				\cline{3-4}
				\cline{5-6}
				
				Isotope  & $J_i^{\pi}\rightarrow J_f^{\pi}$  & G-matrix &  Exp.  & G-matrix  &  Exp. \T\B\\
				
				\hline

				$^{98}$Cd    & $8^+_1\rightarrow 6^+_1$  & 0.005 &  0.0038(4) & 0.81 $\mu$s  & 0.154(16) $\mu$s \T\B\\
				
				$^{100}$Cd & $8^+_1\rightarrow 6^+_1$  & 0.0002  &  0.000042(5)  & 4.33$\times10^{-3}\mu$s  &    0.073 $\mu$s \T\B\\
				& $8^+_1\rightarrow 6^+_2$  & 0.0079  &   0.0061(+33-28) & 10.03 $\mu$s  &  0.413 $\mu$s \T\B\\
				
				$^{102}$Cd &  $8^+_1\rightarrow 6^+_1$  & 0.007  &  0.000017(3) & 97.5$\times10^{-6}\mu$s  & 0.093 $\mu$s \T\B\\
				& $8^+_1\rightarrow 6^+_2$  & 0.004  &   -  & 0.12 $\mu$s  & - \T\B\\
				& $8^+_1\rightarrow 6^+_3$  & 0.0006 & -  & 49.3 $\mu$s  & 0.067 $\mu$s \T\B\\
				
				$^{104}$Cd &  $8^+_1\rightarrow 6^+_1$ & 0.1087  & 0.0015(5) & 3.12$\times10^{-6}\mu$s & {\color{black}8.6$\times10^{-4}\mu$s} \T\B\\
				& $8^+_1\rightarrow 6^+_2$  & 0.0001 &  0.0002(1) & 37.07$\times10^{-3}\mu$s   & {\color{black}11.43$\times10^{-3}\mu$s} \T\B\\
				
				$^{130}$Cd &  $8^+_1\rightarrow 6^+_1$  & 0.006   & - & 1.12 $\mu$s   & 0.220(30) $\mu$s \T\B\\
				\hline		
			\end{tabular}
			\label{8+BE2}
		\end{table}
		
		\subsection{$8^{+}$ isomeric states in Cd chain }
		
\begin{figure}
			\includegraphics[width=81mm]{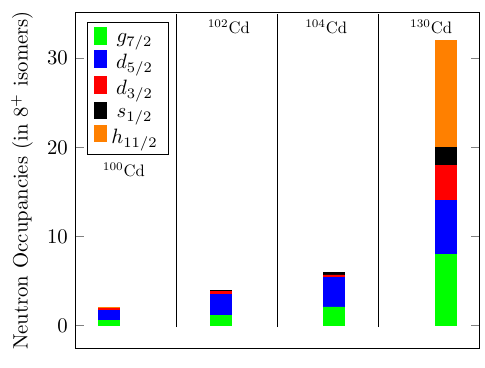}
			\caption{\label{8+isomer} Neutron occupancies in $8^{+}$ isomeric states for $^{100-104,130}$Cd isotopes.}
		\end{figure}

		Many theoretical and experimental investigations have been performed to study isomeric states in the different mass regions of the nuclear chart \cite{Srivastava, Bhoy, Bhoy1,  Srivastava1, Garg, Axel, Wimmer}. The $8^{+}$ isomers have been observed experimentally in $^{98-104,130}$Cd isotopes. In this section, we briefly discussed the $B(E2)$ transitions, neutron occupancies, configurations of the $8^{+}$ isomeric states, and their g-factor values in the Cd chain. The $B(E2)$ transitions and half-lives of $8^+_1$ isomeric states are reported in Table \ref{8+BE2};
  their shell model configurations, 
  and the g-factor values are reported in Table \ref{isomer} and \ref{g-factor}, respectively. The neutron occupancies corresponding to $8^+$ isomeric states are shown in Fig. \ref{8+isomer}.

		\begin{table}
			\caption{Configurations of $8^+_1$ isomeric states in Cd isotopes with the probability of the dominant component of the configuration.}
			\begin{tabular}{cccc}
				\hline
				Isotope	& $J^{\pi}$   & Configuration & Probability (\%)  \\
				\hline
				$^{98}$Cd	& $8^+_1$   & $\pi(p_{1/2}^2g_{9/2}^8)$ & 100 \\
				
				$^{100}$Cd	& $8^{+}_1$  & $\pi(p_{1/2}^2g_{9/2}^8)\otimes\nu(d_{5/2}^2)$ & 55 \\
				
				$^{102}$Cd	& $8^{+}_1$   & $\pi(p_{1/2}^2g_{9/2}^8)\otimes\nu(d_{5/2}^2g_{7/2}^2)$ & 32\\
				
				$^{104}$Cd	&  $8^{+}_1$& $\pi(p_{1/2}^2g_{9/2}^8)\otimes\nu(d_{5/2}^4g_{7/2}^2)$ & 40 \\
				
				$^{130}$Cd	&  $8^{+}_1$  & $\pi(p_{1/2}^2g_{9/2}^8)\otimes\nu(d_{5/2}^4s_{1/2}^2g_{7/2}^2d_{3/2}^4h_{11/2}^{12})$ & 100 \\
				\hline	
				
			\end{tabular}
			\label{isomer}
		\end{table}
		
		In the case of $^{98}$Cd, our result for $B(E2;8^+_1\rightarrow6^+_1)$ transition is showing reasonable agreement with the experimental data. The experimental and calculated $B(E2;8^+_1\rightarrow6^+_1)$ values are 0.0038(4) $e^2b^2$ and 0.005 $e^2b^2$, respectively. As reported in Table \ref{8+BE2}, the small value of $B(E2;8^+_1\rightarrow6^+_1)$ supports $8^{+}_1$ state is an isomeric state with calculated half-life 0.81 $\mu$s.
		
		For $^{100}$Cd, the $8^{+}_1$ state arises because of one pair breaking in $g_{9/2}$ proton orbital and shows proton character. The configuration of $6^{+}_1$ state is $\pi(p_{1/2}^2g_{9/2}^8)_{0^+}\otimes\nu(g_{7/2}^1d_{5/2}^1)_{6^+}$ with 82.4\% probability, which shows predominant neutron character of the $6^{+}_1$ state. But, the configuration of $6^{+}_2$ state is $\pi(p_{1/2}^2g_{9/2}^8)_{6^+}\otimes\nu(d_{5/2}^2)_{0^+}$ with 73.3\% probability. Here, the $6^{+}_2$ state is formed by one proton pair breaking in $g_{9/2}$ orbital. So, the $6^{+}_2$ state has more resemblance to the $8^{+}_1$ isomer in $^{100}$Cd. For this isotope, there are a difference in $B(E2;8^+_1\rightarrow6^+_1)$ values of our calculated result and the experimental data. But, the $B(E2;8^+_1\rightarrow6^+_2)$ transition is comparable with the experimental value. Clark $et$ $al.$ \cite{Clark} reported the experimental branching ratios of $B(E2)$ transitions in $8^{+}_1$ state in $^{100}$Cd. We have used their data to find the partial half-lives of $8^{+}_1$ state as reported in Table \ref{8+BE2}.
		
		\begin{table}
			\caption{Calculated g-factors (in $\mu_N$) for the experimentally observed $8^{+}$ isomeric states in Cd isotopes compared to the experimental data. 
   The shell-model results are evaluated with $g^{\pi}_l=1.0$, $g^{\nu}_l=0.0$, $g^{\pi}_s=3.910$, and $g^{\nu}_s=-2.678$. }
			\begin{tabular}{cccc}
				\hline
				Isotope	&	$J^{\pi}$ 	& G-matrix &	Exp.	\T\B \\
				\hline
				$^{98}$Cd	&	$8^{+}_1$   & 1.323 &  - \T\B \\
				$^{100}$Cd	&	$8^{+}_1$   & 1.269 &  1.24(6) \T\B \\
				$^{102}$Cd	&	$8^{+}_1$   & 1.140 &  1.29(3) \T\B \\
				$^{104}$Cd	&	$8^{+}_1$   & 0.392 & - \T\B \\
				$^{130}$Cd	&	$8^{+}_1$   & 1.323 &  - \T\B \\
				\hline				
			\end{tabular}
				\label{g-factor}
			
		\end{table}
	
		In the case of $^{102,104}$Cd, we have used the experimental data of branching ratios from the Ref. \cite{Lieb, Sukhoruchkin} to calculate the partial half-lives of $8^{+}_1$ state, respectively. For $^{130}$Cd, the calculated half-lives of $8^{+}_1$ isomeric state is 1.12 $\mu$s. Which shows reasonable agreement with the experimental half-life (0.220(30) $\mu$s \cite{NNDC}).
		
		In the vicinity of the doubly magic nuclei $^{132}$Sn, microsecond isomers are studied in the previous work \cite{Pinston}. Scherillo $et$ $al.$ \cite{Scherillo} investigated microsecond isomers in the neutron-rich region of the Cd isotopes using the LOHENGRIN mass spectrometer. But, the expected $8^{+}$ isomeric states were not observed. Our shell-model calculations predict the half-lives of the $8^{+}_1$ states in the ns range for $^{126,128}$Cd. The half-lives with our calculations are much shorter, which supports the non-observation of microsecond isomers in $^{126,128}$Cd isotopes.
		
		In Fig. \ref{8+isomer}, we have shown the occupancy of neutron orbitals corresponding to $8^{+}_1$ isomeric states in $^{100-104,130}$Cd isotopes with our calculations. For $^{100-104}$Cd isotopes, $d_{5/2}$ and $g_{7/2}$ neutron orbitals are dominant corresponding to $8^{+}_1$ isomeric states.

The experimental data of the g-factors for $8^{+}_1$ isomeric states in Cd isotopes are available only for $^{100}$Cd and $^{102}$Cd. 
The $g$-factor can be a measure of the purity of a configuration and the contribution of the component corresponding to a particular state \cite{Bhoy}. 
In the current study, since $^{98}$Cd has only two proton holes in the model space, the $8^+_1$ wave function has the pure  $(\pi0g_{9/2}^{-2})_{J=8}$ aligned configuration.
In the shell-model results, the g-factors of the isomeric $8^+_1$ state are rather constant for $^{98-104,130}$Cd isotopes, except for $^{104}$Cd. It indicates that these states are dominated by the $(\pi0g_{9/2}^{-2})_{J=8}$ configuration.
Concerning the shell-model result of $^{104}$Cd, the $8^+_2$ state locates 110 keV above the $8^+_1$ state, and its g-factor is 1.142, close to the value of $(\pi0g_{9/2}^{-2})_{J=8}$ configuration.

  	\subsection{Energy surface by the Hartree-Fock-Bogoliubov calculation} 
		\label{HFB}
		
		\begin{figure*}[h]
			 
			\includegraphics[width=52mm]{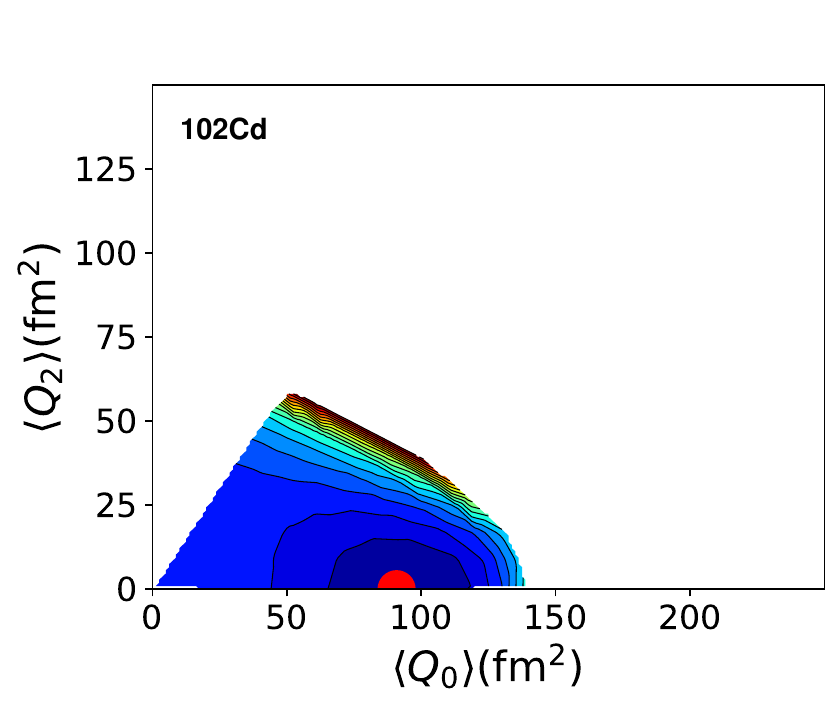}
			\includegraphics[width=60mm]{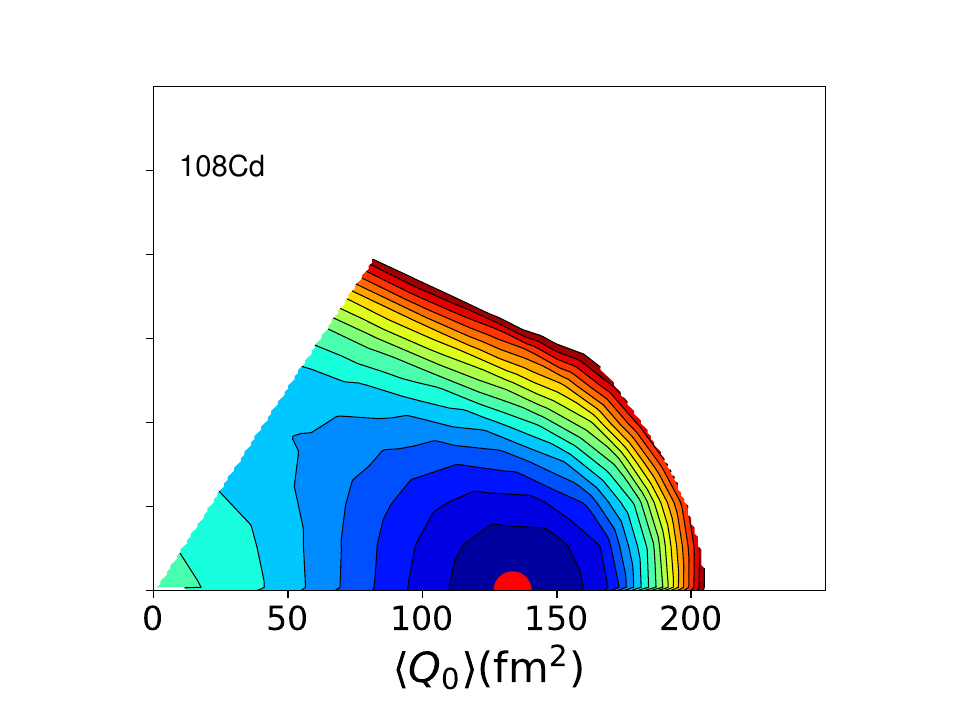}
		  \includegraphics[width=60mm]{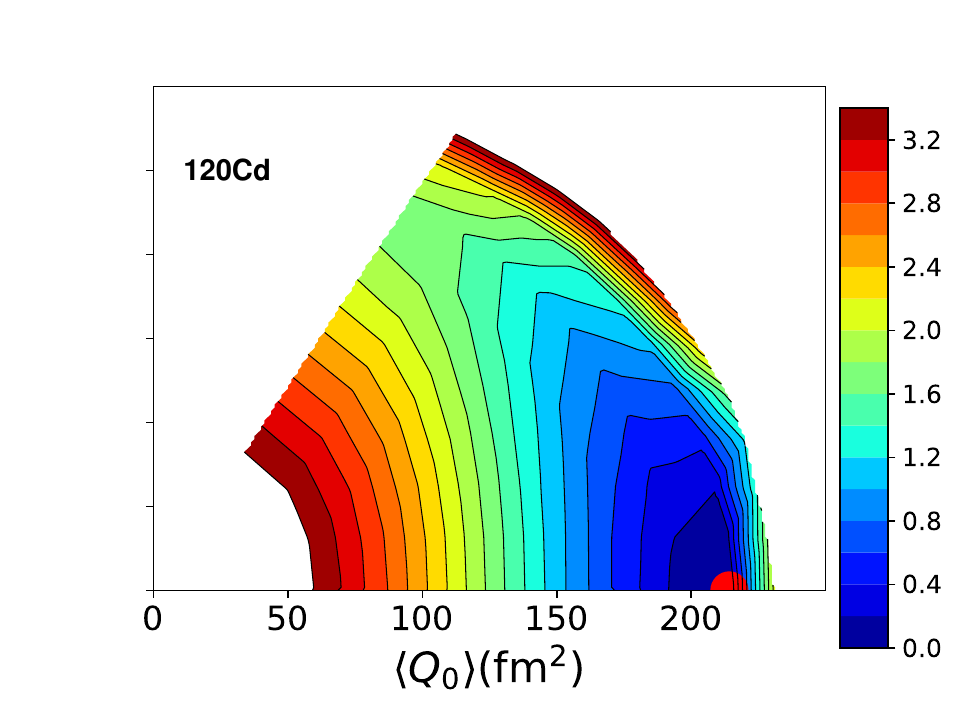} 
			\caption{\label{fig15} Energy surfaces of $^{102,108,120}$Cd isotopes provided by the quadrupole-constrained Hartree-Fock-Bogoliubov calculations. The red circle denotes the energy minimum. The contour interval   is 200 keV. }
		\end{figure*}
		

		We present the energy surfaces of the intrinsic states of Cd isotopes utilizing the Hartree-Fock-Bogoliubov (HFB) method \cite{Shimizu}. In this method, we performed the quadrupole-constrained HFB calculation with variation after number projection employing the same model space and Hamiltonian with the previous sections. The energy surfaces were also shown in recent previous works to discuss the quadrupole collective states and their transitions of Ni, Xe, Ba, and Nd isotopes \cite{Shimizu, Kaneko, Kisyov, Kaneko1}. Figure \ref{fig15} shows the energy surfaces of $^{102,108,120}$Cd as functions of quadrupole moments, $\langle \hat{Q}_{0} \rangle $ and $\langle \hat{Q}_{2} \rangle $. Here, the quadrupole-moment operators are defined as $\hat{Q}_m=r^2Y^{(2)}_m$ with $m=0$ or 2. While the surface of $^{102}$Cd shows small prolate deformation, the deformation grows gradually as the neutron number increases and reaches maximum at $^{120}$Cd.
        On the more neutron-rich side, the deformation gradually decreases. These behaviors are consistent with the fact that the shell-model excitation energies of the yrast levels reach a minimum and the ratio $Ex(4^+_1)/Ex(2^+_1)$ reaches a maximum at $N=70$ as shown in Figs.~\ref{figE2} and \ref{E42}. Since the present shell-model result seriously underestimates the excitation energies around $N=70$, the shell-model predictions overestimate the quadrupole deformation as discussed in Figs.~\ref{BE2fig} and \ref{electric}. Further improvement of the shell-model Hamiltonian is future work.
		
		\section{Summary and Conclusions} \label{section5}
		
		In the present work, we have investigated the low-lying states in $^{98-130}$Cd isotopes. Systematic shell-model calculations have been performed for Cd isotopic chain from $N=50$ to $N=82$ with the G-matrix interaction. From our calculation for $^{98-104}$Cd isotopes with the G-matrix interaction, we can see that the energy gap between $0^{+}_1$ and $2^{+}_1$ state increases as we move from $^{104}$Cd to $^{98}$Cd. This indicates we are approaching the collective vibrational to the closed shell region. The behavior of $4^{+}_1/2^{+}_1$ energy ratio ($R_{4/2}$) has been studied in the Cd chain. We have shown the trend of $B(E2;0^{+}_1\rightarrow 2^{+}_1)$ transition along Cd isotopic chain and tried to explain the asymmetry near $N=62$. We have also calculated the other electromagnetic properties like electric quadrupole moments, $B(M1)$ transitions, and magnetic moments for even $^{98-130}$Cd isotopes. We have studied the behavior of octupole collective $3^{-}_1$ states in Cd isotopes. The properties of experimentally observed $8^{+}$ isomeric states in $^{98-104,130}$Cd isotopes have been investigated using shell-model calculations. Our study also supports the occurrence of E2 isomerism in $8^{+}_1$ states in neutron-deficient Cd isotopes. We have reported the g-factor values for the $8^{+}_1$ isomeric states in Cd isotopes and tried to explain the purity of the configuration of these states. 
		It is important to mention here that  G-matrix interaction fails quantitatively in the mid-shell region, even for the energies of the $2_1^+$ excitations, thus further improvement is needed. The energy surface plot for a few isotopes is also shown to show an increase of deformation in the mid-shell.

		\section{Acknowledgements}

		We acknowledge financial support from MHRD, the Government of India, and SERB (India),
		CRG/2022/005167. We would like to thank the National Supercomputing Mission (NSM) for providing computing resources of ‘PARAM Ganga’ at the Indian Institute of Technology Roorkee, implemented by C-DAC and supported by the Ministry of Electronics and Information Technology (MeitY) and Department of Science and Technology (DST), Government of India. We would also like to thank Professor Larry Zamick for useful discussions during this work.   P.C.S. acknowledges the hospitality
		extended to him during his stay at the Center for Computational Sciences, University of Tsukuba, Japan.
        We acknowledge N. Smirnova for the $G$-matrix shell-model interaction and Y. Utsuno for fruitful discussions.
        N.S. acknowledges the support of ``Program for promoting researches on the supercomputer Fugaku'', MEXT, Japan (JPMXP1020230411) and the MCRP program of the Center for Computational Sciences, University of Tsukuba (wo22i002).
		

	\end{document}